\documentclass{llncs}

\usepackage{times}
\usepackage{amsmath, amssymb, mathrsfs}  % for some math symbols, if needed
\usepackage{graphicx}                    % pictures
\usepackage{color}
\usepackage[shortend,ruled,algo2e]{algorithm2e} % algo2e = use \begin{algorithm2e}
\usepackage[section]{algorithm}                  % for algorithm floats
\usepackage{subfigure}                   % for multiple figures
\usepackage{url}                         % to format URLs
\usepackage{multirow}                    % multirow and multicolumn support
\usepackage{array}                       % flexible tables
\usepackage{tikz}                        % diagrams
\usepackage{verbatim}
\usepackage{hyperref}
\usepackage{rotating}                    % sidewaysfigure
\usepackage{mathabx}                     % \notequiv
\usepackage{stmaryrd}

\usepackage{arydshln}                    % dashed lines in tables

\numberwithin{equation}{section}
\numberwithin{figure}{section}
\numberwithin{table}{section}

\newcommand{\defterm}[1]{\emph{#1}}      % definition of a term
\newcommand{\todo}[1]{{}} %{{\sf TODO: #1}}
\newcommand{\tuple}[1]{\langle #1 \rangle} % tuple (in mathmode)
\newcommand{\class}[1]{\mathscr{#1}}     % class (in mathmode)
\newcommand{\fml}[1]{{\mathcal{#1}}}     % propositional formulas (in mathmode)
\newcommand{\SAT}{\mathsf{SAT}}
\newcommand{\UNSAT}{\mathsf{UNSAT}}

\newcommand{\MUS}{\mathsf{MUS}}
\newcommand{\MSS}{\mathsf{MSS}}
\newcommand{\coMSS}{\mathsf{coMSS}}
\newcommand{\MES}{\mathsf{MES}}
\newcommand{\MNS}{\mathsf{MNS}}
\newcommand{\coMNS}{\mathsf{coMNS}}

\newcommand{\LMUS}{\mathsf{LMUS}}
\newcommand{\LMSS}{\mathsf{LMSS}}
\newcommand{\coLMSS}{\mathsf{coLMSS}}
\newcommand{\LMES}{\mathsf{LMES}}
\newcommand{\LMNS}{\mathsf{LMNS}}
\newcommand{\coLMNS}{\mathsf{coLMNS}}

\newcommand{\PP}{\mathsf{P}}

\newcommand{\lcons}{\vDash}                     % logical consequence
\newcommand{\N}{\mathbb{N}}                      % set of natural numbers

\definecolor{midgrey}{rgb}{0.5,0.5,0.5}
\definecolor{darkred}{rgb}{0.7,0.1,0.1}

\begin{document}

\title{Generalizing Redundancy in Propositional Logic:\\Foundations and Hitting Sets Duality}
\author
{
Anton Belov \inst{1}
\and
Joao Marques-Silva\inst{1,2}
\!\!%
\color{white}\thanks{
  This work is partially supported by SFI grant BEACON (09/IN.1/I2618),
  and by FCT grants ATTEST (CMU-PT/ELE/0009/2009) and POLARIS
  (PTDC/EIA-CCO/123051/2010).}
}

\institute{
CASL, University College Dublin, Ireland
\and
IST/INESC-ID, Lisbon, Portugal
}

\maketitle

%------------------------------------------------------------------------------%
%
% Generalized redundancy (arXiv version), abstract
%
\begin{abstract}
Detection and elimination of redundant clauses from propositional formulas 
in Conjunctive Normal Form (CNF) is a fundamental problem with
numerous application domains, including AI, and has been the subject of
extensive research.
Moreover, a number of recent applications motivated various extensions
of this problem. For example, unsatisfiable formulas partitioned into
disjoint subsets of clauses (so-called \defterm{groups}) often need to
be simplified by removing redundant groups, or may contain redundant
\emph{variables}, rather than clauses.
In this report we present a generalized theoretical framework of 
\emph{labelled CNF formulas} that unifies various extensions of the 
redundancy detection and removal problem and allows to derive a number 
of results that subsume and extend previous work. 
The follow-up reports contain a number of additional theoretical results 
and algorithms for various computational problems in the context of the
proposed framework.
\end{abstract}

%%% Local Variables:
%%% mode: pdflatex
%%% TeX-master: "paper"
%%% End:

%
% Generalized redundancy (arXiv version)
%

\section{Introduction}\label{sec:intro}

Propositional logic formulas in Conjunctive Normal Form (CNF) often
have redundant clauses.
In some contexts, redundancy is desirable. For example, the
identification of redundant clauses is a hallmark of modern SAT
solvers~\cite{jpms-sathbk09}.
In other contexts, redundancy is undesirable. For example, elimination
of redundant clauses is useful in simplifying knowledge
bases~\cite{liberatore-aij05}.
A special case of redundancy deals with unsatisfiable subformulas,
since the identification of Minimal Unsatisfiable Subformulas (MUSes)
finds a wide range of practical applications.

Redundancy in logic has been extensively studied in the recent
past~\cite{roussel-aaai00,liberatore-aij05,sais-cpaior07,liberatore-aij08a,liberatore-aij08b},
and includes complexity characterizations of different computational
problems. Similarly, the specific case of unsatisfiable subformulas
has also been extensively
studied~\cite{kullmann-cc00,kullmann-sat06,kullmann-fi11b}.
Computational problems of interest include computing a minimal
unsatisfiable subformula, or enumerating them all, and computing an
irredundant (or minimal equivalent) subformula, or enumerating them
all. Some of these problems have been studied in detail for the case
where minimality is expressed in terms of clauses.
Moreover, and also for the case where minimality is expressed in terms
of clauses, well-known hitting set properties relating minimal
unsatisfiable and maximal satisfiable subformulas have been developed 
for unsatisfiable
formulas~\cite{reiter-aij87,lozinskii-jetai03,kullmann-cc00}. 
Recently, this work has been extended to the case of
satisfiable formulas~\cite{kullmann-fi11b}.

Motivated by practical applications, the extraction of MUSes has
recently been generalized to groups of (related)
clauses~\cite{kas-jar08,nadel-fmcad10}, and to
variables~\cite{chen-06,chen-08,desrosiers-dam08,hertz-jco09}. In many 
settings~\cite{kas-jar08,nadel-fmcad10}, it is important to aggregate
related clauses (as {\em groups} of clauses). In these cases, MUSes
need to be expressed in terms of groups of clauses and not in terms of
individual clauses.
Clearly, MUS problems over groups of clauses or over variables can be
extended to the more general case of redundancy removal. For example,
one may want to compute a subformula that has no redundant variables,
or a subformula that has no redundant groups of clauses.
Also relevant are enumeration problems for unsatisfiability and
redundancy problems when these problems are expressed in terms of
variables or groups of clauses. For example, one may want to enumerate
all the variable MUSes of a formula, or all the irredundant
subformulas when a problem is represented as groups of (related)
clauses. 

The main objective of this report is to develop a theoretical framework
that provides a unified approach for tackling redundancy problems in
CNF formulas, and includes unsatisfiable formulas as a special case.
This framework enables the generalization of known theoretical
results, but also serves to highlight how existing algorithms for
different computational problems can be adapted and
extended~\cite{msl-sat11,bms-fmcad11,bimms-sat12}.
The framework is based on the concept of {\em labelled CNF} formula, 
where labels are used to associate individual clauses
of a CNF formula
with disjoint groups of clauses, or
with variables, or with literals, or even with arbitrary intersecting
groups of clauses.
By extending to the labelled CNF setting the standard definitions of 
MUSes and MSSes over clauses, the report shows that well-known
properties of hitting set
duality~\cite{reiter-aij87,kullmann-cc00,lozinskii-jetai03} also hold
for the general case of unsatisfiable labelled CNF formulas, and so
hold for MUS and MSS problems over variables, literals or arbitrary
groups of clauses. 
More interestingly, these results also hold for redundancy removal
problems for satisfiable formulas, when defined over clauses,
variables, or groups of clauses.
The immediate consequences of these results include the ability to
enumerate MSSes and MUSes of labelled CNF formulas, their extensions
to the redundancy removal case, but also the ability to generalize
existing MUS extraction algorithms.
A detailed description of the report's contributions is included 
in Section~\ref{sec:mot} and summarized in Table~\ref{tab:contrib}.

%%% Local Variables:
%%% mode: pdflatex
%%% TeX-master: "paper"
%%% End:

%
% Generalized redundancy (arXiv version)
%

\section{Background and Motivation}\label{sec:mot}

We focus on formulas in CNF (\defterm{formulas}, from hence on), which we treat as finite
multi-sets of clauses. We assume that clauses do not contain duplicate variables.
Given a formula $\fml{F}$ we denote the set of variables that occur in $\fml{F}$ by $Var(\fml{F})$,
and the set of variables that occur in a clause $c \in \fml{F}$ by $Var(c)$. An \defterm{assignment}
$\tau$ for $\fml{F}$ is a map $\tau: Var(\fml{F}) \to \{0,1\}$. Assignments are extended to clauses 
and formulas according to the semantics of classical propositional logic. 
%By $Unsat(\fml{F},\tau)$ we denote the set of clauses of $\fml{F}$ falsified by $\tau$.
If $\tau(\fml{F}) = 1$, then $\tau$ is a \defterm{model} of $\fml{F}$. If a formula $\fml{F}$ has 
(resp. does not have) a model, then $\fml{F}$ is \defterm{satisfiable} (resp. \defterm{unsatisfiable}). 
By $\SAT$ (resp. $\UNSAT$) we denote the set of all satisfiable (resp. unsatisfiable) CNF formulas.
Formula $\fml{F}_1$ \defterm{implies} formula $\fml{F}_2$  ($\fml{F}_1 \lcons \fml{F}_2$) if every model of 
$\fml{F}_1$ is a model of $\fml{F}_2$. $\fml{F}_1$ is \defterm{equivalent} to $\fml{F}_2$ 
($\fml{F}_1 \equiv \fml{F}_2$) if they have the same set of models. 
A clause $c \in \fml{F}$ is \defterm{redundant} in $\fml{F}$ if $\fml{F} \setminus \{ c \} \equiv \fml{F}$,
or, equivalently, $\fml{F} \setminus \{ c \} \lcons \{ c \}$. Formulas with (resp. without) redundant 
clauses are called \defterm{redundant} (resp. \defterm{irredundant}).

The majority of the research on redundancy in propositional logic addresses unsatisfiable CNF formulas.
Irredundant unsatisfiable formulas are called \defterm{minimally unsatisfiable (MU)}. Explicitly, a
formula $\fml{F}$ is MU if (i)~$\fml{F} \in \UNSAT$, and (ii)~for any
clause $c \in \fml{F}$,  $\fml{F} \setminus \{c\} \in \SAT$. A subformula $\fml{F}' \subseteq \fml{F}$ 
is a \defterm{minimally unsatisfiable subformula (MUS)} of $\fml{F}$ if $\fml{F}'$ is minimally 
unsatisfiable. %i.e. $\fml{F}' \in \UNSAT$, and $\forall c \in \fml{F}'$, $\fml{F}' \setminus \{ c \} \in \SAT$. 
The set of all MUSes of $\fml{F}$ is denoted by $\MUS(\fml{F})$ --- in general, 
a given unsatisfiable $\fml{F}$ may have more than one MUS. MUSes are of interest for a number
of reasons, and have been on the radar of AI community for a long time. For example, in early
work of Reiter on model-based diagnosis \cite{reiter-aij87}, MUSes, under the name of 
\defterm{minimal conflict sets}, are used in computation of a faulty set of components of
mis-behaving systems. More recently, MUSes find numerous applications in formal verification
of hardware and software systems, product configuration, etc. --- see \cite{jpms-ismvl10} for
concrete examples. Motivated by several applications, minimal unsatisfiability and related concepts 
have been extended to CNF formulas where clauses are partitioned into disjoint sets called 
\defterm{groups}~\cite{kas-jar08,nadel-fmcad10}.
\begin{definition}[Group-Oriented MUS] Given an explicitly partitioned
unsatisfiable CNF formula $\fml{F} = \fml{G}_0 \cup \dots \cup \fml{G}_n$, a 
\defterm{group oriented MUS} (or, \defterm{group-MUS}) of $\fml{F}$ is a set 
of groups $\{ \fml{G}_{i_1}, \dots,\fml{G}_{i_k} \}$, $i_j > 0$, such that 
$\fml{F}' = \fml{G}_0 \cup \fml{G}_{i_1} \cup \dots \cup \fml{G}_{i_k} \in \UNSAT$, and
for every $1\le j\le k$, $\fml{F}'\setminus \fml{G}_{i_j} \in \SAT$.
\end{definition}
Note the special role of group $\fml{G}_0$ (\defterm{group-0}) --- this group consists of 
``background'' clauses that are included in every group-MUS; because of group-0 a group-MUS,
as opposed to MUS, can be empty.
In addition to clauses and groups of clauses, minimal unsatisfiability has been defined 
and analysed in terms of the \emph{variables} of the formula \cite{chen-06,hertz-jco09}.
Given a CNF formula $\fml{F}$, and $V \subseteq Var(\fml{F})$, the subformula of $\fml{F}$ 
\defterm{induced} by $V$ is the formula $\fml{F}|_V = \{ c \in \fml{F}\ |\ Var(c) \subseteq V \}$.
Then, $\fml{F}$ is \defterm{variable minimally unsatisfiable (VMU)} if $\fml{F} \in \UNSAT$, and for any
$V \subset Var(\fml{F})$, $\fml{F}|_V \in \SAT$, i.e. no variable can be removed from 
the formula without making it satisfiable. Here ``removal of a variable'' means removal of all
clauses that have this variable. \defterm{Variable MUSes (VMUSes)} are defined accordingly: 
$V \subseteq Var(\fml{F})$ is a VMUS of $\fml{F}$ if $\fml{F}|_V$ is VMU. In \cite{bimms-sat12}
variable minimal unsatisfiability has been extended in a number of ways akin to the extension
of MUSes with group-MUSes.

A notion dual to minimal unsatisfiability is that of maximal satisfiability: a subformula 
$\fml{F}' \subseteq \fml{F}$ is a \defterm{maximally satisfiable subformula (MSS)} of 
$\fml{F}$ if $\fml{F}' \in \SAT$ and $\forall c \in \fml{F} \setminus \fml{F}'$, 
$\fml{F}' \cup \{ c \} \in \UNSAT$. The set of MSSes of a CNF formula $\fml{F}$ is 
denoted by $\MSS(\fml{F})$. MSSes are also of much interest in the context of AI.
For once, given that an MSS constitutes a maximally consistent part of an
inconsistent (i.e. unsatisfiable) formula, MSSes can be used for reasoning in 
the presence of inconsistency --- see \cite{lozinskii-jetai03} for an example of 
an MSS-based framework for reasoning with inconsistent knowledge. Furthermore, an MSS of maximum cardinality constitutes a set of clauses satisfied by a solution to the Maximum Satisfiability (MaxSAT) problem: given a formula $\fml{F}$ find an assignment that satisfies the maximum number of clauses of $\fml{F}$.

Given an MSS $\fml{S}$ of $\fml{F}$, one may also consider a subformula 
$\fml{F} \setminus \fml{S}$ of $\fml{F}$ --- such subformula is called
a \defterm{co-MSS} of $\fml{F}$, and the set of all co-MSSes of $\fml{F}$ is 
denoted by $\coMSS(\fml{F})$. Note that when $\fml{F} \in \UNSAT$, a co-MSS of $\fml{F}$ 
is a minimal subformula of $\fml{F}$, removal of which from $\fml{F}$ will 
regain its satisfiability. Thus, for example, in the context of Reiter's model-based
diagnosis framework \cite{reiter-aij87}, co-MSSes constitute the minimal set of
components of the faulty system that must be removed to restore its correct behaviour,
i.e. the \defterm{minimal diagnosis}. For a similar reason, in \cite{kas-jar08} the 
authors refer to co-MSSes are \defterm{minimal correction subsets (MCSes)}.

The MUSes, MSSes and co-MSSes of a given unsatisfiable formula $\fml{F}$ are connected
via so-called \defterm{hitting sets duality theorem}. This theorem has been proved and
re-proved on a number of occasions, starting with \cite{reiter-aij87}, and later in 
\cite{lozinskii-jetai03,bruni-dam03,stuckey-padl05,kas-jar08}. The connection is expressed 
in terms of \defterm{irreducible hitting sets}.
\begin{definition}[(Irreducible) Hitting Set]\label{def:hs}
Let $\class{S}$ be a collection of arbitrary sets. A set $H$ is called a \defterm{hitting set} 
of $\class{S}$ if for all $S \in \class{S}$, $H \cap S \ne \emptyset$. A hitting set $H$
is \defterm{irreducible}, if no $H' \subset H$ is a hitting set of $\class{S}$.
\end{definition}
Then, the hitting set duality theorem states that every MUS of a formula $\fml{F}$ is
an irreducible hitting set of the set of co-MSSes of $\fml{F}$, and vice versa.
\begin{theorem}[cf. \cite{reiter-aij87,lozinskii-jetai03,bruni-dam03,stuckey-padl05}]\label{th:duality}
For any unsatisfiable CNF formula $\fml{F}$: $(i)$ formula $\fml{M}$ is a co-MSS of $\fml{F}$ 
if and only if $\fml{M}$ is an irreducible hitting set of $\MUS(\fml{F})$;
$(ii)$ formula $\fml{U}$ is an MUS of $\fml{F}$ if and only if $\fml{U}$ is an irreducible 
hitting set of $\coMSS(\fml{F})$.
\end{theorem}
Besides exposing an interesting connection between the various subformulas of CNF formulas,
hitting set duality is used in algorithms for computation of the set of \emph{all} MUSes of 
CNF formulas --- see, for example, \cite{stuckey-padl05,kas-jar08}.

The case of redundancy in satisfiable CNF formulas has also been analysed extensively, for example
in \cite{liberatore-aij05,kullmann-sat06,kullmann-fi11b}. Here the first object of interest is
a subformula of a CNF formula $\fml{F}$ that is irredundant and equivalent to $\fml{F}$ --- such
subformulas are called \defterm{minimal equivalent subformulas (MESes)}: 
a subformula $\fml{F}' \subseteq \fml{F}$ is an MES of $\fml{F}$ if 
$\fml{F}' \equiv \fml{F}$, and $\forall c \in \fml{F}'$, $\fml{F}' \setminus \{ c \} \notequiv \fml{F}$.
The set of all MESes of $\fml{F}$ is denoted by $\MES(\fml{F})$. A number of efficient
algorithms for computation of MESes have recently been proposed in  \cite{bjlms-cp12-sub}.
The dual notion is that of a \defterm{maximal non-equivalent subformula (MNS)}: a subformula 
$\fml{F}' \subseteq \fml{F}$ is an MNS of $\fml{F}$ if $\fml{F}' \notequiv \fml{F}$ and 
$\forall c \in \fml{F} \setminus \fml{F}'$, $\fml{F}' \cup \{ c \} \equiv \fml{F}$. The set of 
MNSes  of a CNF formula $\fml{F}$ is denoted by $\MNS(\fml{F})$. Finally, a subformula of
$\fml{F}$ that is a complement of some MNS of $\fml{F}$ is called a \defterm{co-MNS} of $\fml{F}$,
and the set of all co-MNSes of $\fml{F}$ is denoted by $\coMNS(\fml{F})$. Note that, as opposed
to the case of unsatisfiable formulas, to our knowledge no extensions of MESes and related concepts, 
to groups of clauses or to the variables of CNF formulas have been proposed.

%% jpms: contributions in a separate file !!
%
% Generalized redundancy (arXiv version)
%

%\subsection{Overview of Contributions} \label{ssec:contrib}

%%From MUS to MES. From clauses to intersecting groups. Plus,
%%wire-MUSes for circuits.
%% \anote{Matrix goes here !}

\begin{table}[t]
\caption{Summary of existing work on redundancy in CNF formulas. The framework of 
\defterm{labelled CNF formulas} proposed in this report allows to ``cover'' all 
the empty entries.}\label{tab:contrib}
\centering
\begin{tabular}{| c | >{\centering\arraybackslash} p{1.25cm} || >{\centering\arraybackslash} p{2.35cm} | >{\centering\arraybackslash} p{2.35cm} | >{\centering\arraybackslash} p{2.35cm} | } \hline
\multicolumn{2}{|c||}{Problem} & Clauses & Groups & Variables \\ \hline\hline
\multicolumn{2}{|c||}{MUS/MSS/coMSS} & \cite{mazure-ictai08,kullmann-sathbk09,hertz-jco09,jpms-ismvl10} & \cite{kas-jar08,nadel-fmcad10} & \cite{chen-06,hertz-jco09,bimms-sat12} \\ \hline  %({\em overviews})
\multicolumn{2}{|c||}{MES/MNS/coMNS} & \cite{liberatore-aij05,kullmann-sat06,kullmann-fi11b} &  & \\ \hline
\multirow{2}{*}{Hitting Set Theorem} & UNSAT & \cite{reiter-aij87,lozinskii-jetai03,bruni-dam03,kullmann-dam03,stuckey-padl05} & & \\ \cline{2-5}
                                  & SAT & \cite{kullmann-fi11b} & & \\ \hline
\multicolumn{2}{|c||}{MaxSAT (algorithms)} & \cite{manya-sathbk09,ansotegui-aaai10,hmms-aaai11} & \cite{hmms-caai12} & \\ %\cline{2-5}  %({\em overviews})
%\multirow{2}{*}{MaxSAT Algorithms} & UNSAT & \cite{manya-sathbk09,ansotegui-aaai10,hmms-aaai11} & \cite{hmms-caai12} & \\ \cline{2-5}  %({\em overviews})
%                                   & SAT   & & & \\ \hline
\hline
\end{tabular}
\end{table}

Table~\ref{tab:contrib} summarizes existing work on redundancy over
clauses, groups of clauses and variables. A number of concrete
problems and properties can be considered, namely minimal
unsatisfiability, irredundant (or minimal equivalent) subformulas,
hitting set duality theorem and maximum satisfiability. The table
shows references for overviews or key references for each topic.
In the next section we describe a framework of so-called 
\defterm{labelled CNF formulas}. This framework serves to generalize 
all of the existing work described above, and, in particular, 
allows to ``cover'' all of the empty entries in the table. We demonstrate
the usefulness of the framework by deriving a generalized version of 
the hitting set duality theorem. As a by-product we extend the recent 
results on irredundant formulas for the case of satisfiable 
formulas~\cite{kullmann-fi11b}. In addition to the problems shown in 
Table~\ref{tab:contrib}, the framework of labelled CNFs allows addressing redundancy
problems over literals, wire-MUSes for Boolean circuits~\cite{bms-sat11}, and 
\defterm{interesting variables MUS} problem~\cite{bimms-sat12}.

%------------------------------------------------------------------------------%

%%% Local Variables:
%%% mode: pdflatex
%%% TeX-master: "paper"
%%% End:

%
% Generalized redundancy (arXiv version)
%

\section{Generalized Redundancy}\label{sec:genr}

\subsection{Labelled CNF Formulas}\label{ssec:lcnf}

The key observation that motivates the development of the labelled CNF framework
is that in all cases described in Section~\ref{sec:mot} below, the redundancy in a CNF 
formula $\fml{F}$ can be analyzed in terms of \emph{possibly intersecting} (i.e. 
not necessarily disjoint) subsets of clauses of $\fml{F}$. An additional feature 
of some of the cases, for example group-MUS, is the presence of the background, or 
group-0, clauses. We capture the semantics of the intersecting and the background 
subsets of clauses in the following way.
\begin{definition}[Labelled CNF Formula]\label{def:lcnf}
Let $Lbl$ be a non-empty set of \defterm{clause labels}. A 
\defterm{labelled CNF (LCNF) formula} $\Phi$ is a tuple $\tuple{\fml{F}, \lambda}$, where 
$\fml{F}$ is a CNF formula, and $\lambda: \fml{F} \to 2^{Lbl}$ is a (total) 
\defterm{labelling function} such that for all $c \in \fml{F}$, $\lambda(c)$ is finite.
\end{definition}
We refer to the formula $\fml{F}$ as a \defterm{CNF part} of $\Phi$, and denote it 
by $\fml{F}_\Phi$. The labelling 
function $\lambda$ of $\Phi$ is denoted by $\lambda_\Phi$. The set of labels 
$\lambda_\Phi(c)$ for $c \in \fml{F}_\Phi$ is referred to as a 
\defterm{set of clause labels of $c$ in $\Phi$}. For $l \in Lbl$, we refer to the
set of clauses $\fml{F}^l_\Phi = \{ c \in \fml{F}_\Phi\ |\ l \in \lambda_\Phi(c) \}$ as 
the \defterm{set of clauses labelled with $l$}. The role of labels in LCNF formulas 
is to group the clauses of the CNF part into subsets --- these subsets can be disjoint, 
as, for example, in group-CNF context \cite{kas-jar08,nadel-fmcad10}, or intersecting, 
as in the context of variable-MUS problem \cite{chen-06,hertz-jco09}.
By $\fml{F}^\emptyset_\Phi$ we denote the set $\{ c \in \fml{F}_\Phi\ |\ \lambda_\Phi(c) = \emptyset \}$ 
of \defterm{unlabelled clauses}. These clauses play the role of group-0 clauses in 
group-CNFs, or uninteresting variables in the extensions of variable-MUS problem \cite{bimms-sat12}.
The subscripts for the CNF part and the labelling function of $\Phi$ may be omitted 
when $\Phi$ is understood from the context. With a slight abuse of notation, 
by $\lambda(\Phi)$ we denote the set of \defterm{active labels} of $\Phi$, that is the 
set $\bigcup_{c \in \fml{F}_\Phi} \lambda(c)$. Note that $\lambda(\Phi)$ is finite, and
may be empty.
Some natural examples of labelling functions and labelled CNFs will be given shortly.
The (un)satisfiability, models, and all related concepts of propositional logic are 
defined for labelled CNFs with respect to their CNF part. For example, $\Phi$ is 
unsatisfiable ($\Phi \in \UNSAT$), if $\fml{F}_\Phi \in \UNSAT$.
\begin{definition}[Induced subformula]\label{def:ind}
Let $\Phi = \tuple{\fml{F}, \lambda}$ be a labelled CNF formula, and let 
$L \subseteq \lambda(\Phi)$. Then, the subformula of $\Phi$
\defterm{induced by $L$}, is a labelled CNF formula 
$\Phi|_L = \tuple{\fml{F}|_L, \lambda}$, where 
$\fml{F}|_L = \{ c \in \fml{F}\ |\ \lambda(c) \subseteq L \}$.
\end{definition}
In other words, $\Phi|_L$ has the same labelling function $\lambda$ as $\Phi$, however 
the CNF part of $\Phi|_L$ contains only those labelled clauses of $\fml{F}$ \emph{all} 
of whose labels are included in $L$ \emph{and} all the unlabelled clauses $\fml{F}$, i.e. 
$\lambda(\Phi|_L) \subseteq L$. 
Alternatively, any 
clause that has some label outside of $L$ is removed from $\fml{F}$. Thus, it will
be convenient to speak of an operation of \defterm{removal} of a label from 
$\Phi = \tuple{\fml{F}, \lambda}$. Let $l \in \lambda(\Phi)$ be any (active) 
label, then, the LCNF formula $\tuple{\fml{F} \setminus \fml{F}^l, \lambda}$ will be 
said to be obtained by the \defterm{removal of label $l$ from $\Phi$}. 
Note that Definition~\ref{def:ind} implies that for any $L \subset \lambda(\Phi)$ 
(note the strict inclusion), we have $\fml{F}_{\Phi|_L} \subset \fml{F}_{\Phi}$.
Also, note that it is possible that $\lambda(\Phi|_L) \subset L$ --- for example, 
if for some $l \in \lambda(\Phi) \setminus L$, and some $l' \in L$, 
$\fml{F}^{l'} \subseteq \fml{F}^l$, then $l' \notin \lambda(\Phi|_L)$.

\begin{example}\label{ex:1}
Let $Lbl = \N$, and let $\Phi = \tuple{ \{ c_1, \dots, c_8 \}, \lambda}$ with the
clauses $c_i$ and the labelling function $\lambda$ defined as follows (the sets of 
clause labels are shown as subscripts).
\begin{align*}
c_1 &= (\neg y)_{\{1\}}         & c_3 &= (z \lor t)_{\{1\}}   & c_5 &= (x \lor y \lor z)_\emptyset & c_7 &= (\neg y \lor t)_{\{3\}} \\
c_2 &= (y \lor \neg t)_{\{1\}}  & c_4 &= (\neg x)_{\{1,2\}}   & c_6 &= (\neg x \lor y)_{\{2,3\}}  & c_8 &= (\neg t)_{\{4\}}
\end{align*}
The set of active labels of $\Phi$ is $\lambda(\Phi) = \{ 1, 2, 3, 4 \}$. $\Phi$ is satisfiable,
with the (only) model $\{ \neg x, \neg y, z, \neg t \}$. The subformula of $\Phi$ induced by the
set of labels $L = \{ 2, 3, 4 \}$ is $\Phi|_L = \tuple{ \{ c_5, \dots, c_8 \}, \lambda}$. 
Additional examples of induced subformulas are $\Phi|_{\{1,4\}} = \tuple{ \{ c_1, c_2, c_3, c_5, c_8 \}, \lambda}$
and $\Phi|_\emptyset = \tuple{ \{ c_5 \}, \lambda}$.
\end{example}

In the context of redundancy removal in CNF formulas, we speak of redundant clauses,
and the basic, atomic, operation on CNF formulas consists of a removal of a single 
clause from the formula. For the general case of labelled CNF formulas the operation
of removal of a single clause is not permitted --- instead, the atomic modification
to labelled CNFs is a removal of a single (active) \emph{label}, that is \emph{all}
clauses in the CNF part of the formula that are labelled with this label. This is
an {\em essential} point of the framework proposed in this report. In fact, when we speak
of (proper) subformulas of labelled CNF formulas, we \emph{always} mean ``subformulas 
obtained by removal of labels'', or to be precise: $\Phi'$ is a subformula of $\Phi$, 
if $\Phi' = \Phi|_L$ for some $L \subseteq \lambda(\Phi)$. 
When the inclusion is strict, i.e. $L \subset \lambda(\Phi)$, $\Phi'$ is a \emph{proper}
subformula of $\Phi$. We will use set notation to denote subformula relation, e.g.
$\Phi' \subset \Phi$. Note that all subformulas of $\Phi$ have the same set of 
unlabelled clauses. Finally, we point out that while $\Phi' \subseteq \Phi$ implies
$\fml{F}_{\Phi'} \subseteq \fml{F}_{\Phi}$, the fact that $\fml{F}' \subseteq \fml{F}$ does 
necessarily imply $\tuple{\fml{F}', \lambda} \subseteq \tuple{\fml{F}, \lambda}$ --- 
again, because removal of a single clause is, in general, not allowed in LCNFs.

\subsection{Redundancy in Labelled CNFs}\label{ssec:lred}

It is not difficult to see that, similar to the case of (plain) CNF, removal of 
labels from labelled CNF formula can never \emph{reduce} the set of models of
the formulas, that is, when $\Phi'$ is a subformula of $\Phi$, we always have
$\Phi \lcons \Phi'$. However, as with CNFs, removal of some labels from $\Phi$, 
might not affect the set of models of $\Phi$ at all --- such labels are then 
\emph{redundant}, i.e. all clauses that are labelled with such labels can be 
removed from the formula while preserving the logical equivalence.
\begin{definition}[Redundant label; Redundant LCNF]\label{def:red}
Let $\Phi = \tuple{\fml{F}, \lambda}$ be a labelled CNF formula. A label $l \in \lambda(\Phi)$
is \defterm{redundant} in $\Phi$ if $\Phi|_{\lambda(\Phi) \setminus \{ l \}} \equiv \Phi$.
A formula $\Phi$ is \defterm{redundant} if $\lambda(\Phi)$ contains redundant labels.
\end{definition}
Alternatively, a label $l \in \lambda(\Phi)$ is redundant in $\Phi = \tuple{\fml{F}, \lambda}$ if
$(\fml{F} \setminus \fml{F}^l) \lcons \fml{F}^l$. 
An irredundant LCNF has the property that
the removal of any label from it extends the set of its models --- when the formula is unsatisfiable,
this means that the removal of any label makes it satisfiable, i.e. it is \emph{minimally} unsatisfiable.
%For completeness, we give a definition a minimal unsatisfiability for labelled CNFs.
%
\begin{definition}[Minimally Unsatifiable LCNF]\label{def:lmu}
A labelled CNF formula $\Phi = \tuple{\fml{F}, \lambda}$ is 
\defterm{minimally unsatisfiable} if $\Phi \in \UNSAT$, and for any $L \subset \lambda(\Phi)$,
$\Phi|_L \in \SAT$.
\end{definition}
The following example demonstrates a number of natural definitions of labelling functions
under which redundant labels capture some well-known notions of redundancy (cf. Section~\ref{sec:mot}).
\begin{example}\label{ex:2}
Let $\fml{F}$ be any CNF formula.
\begin{itemize}
\item[(i)] Take $\lambda$ to be such that each clause of $\fml{F}$ is labelled with a 
single distinct label. Then a label $l$ is redundant in $\Phi = \tuple{\fml{F},\lambda}$ if and only 
if the (only) clause labelled with $l$ is redundant, in the plain CNF sense, in $\fml{F}$. 
\item[(ii)] Take $\lambda$ to be such that each clause of $\fml{F}$ is either labelled 
with a single, but not necessarily distinct label, or unlabelled. Then a label $l$ is 
redundant in $\Phi = \tuple{\fml{F},\lambda}$ if and only if the set of clauses 
$\fml{F}^l$ is redundant, and so we capture the semantics of redundant groups in the 
group-CNF formulas. The unlabelled clauses $\fml{F}^\emptyset$ correspond to group-0.
\item[(iii)] Take $Lbl = Var(\fml{F})$, and $\lambda(c) = Var(c)$ for each $c \in \fml{F}$. 
Then, a label $v$ is redundant in $\Phi = \tuple{\fml{F},\lambda}$ if and only if the variable $v$
is redundant in $\fml{F}$. Thus, when $\Phi$ is minimally unsatisfiable, $\fml{F}$ is variable
minimally unsatisfiable (VMU).
%, and so we recover the semantics of variable-MUSes.
\end{itemize}
\end{example}

As with the case of CNF, by iteratively removing redundant labels from LCNF $\Phi$ we 
can obtain a subformula $\Phi'$ of $\Phi$ that is equivalent to $\Phi$ and irredundant. Thus,
the subformula $\Phi'$ is a labelled CNF analog of an MES for (plain) CNF formulas (cf.
Section~\ref{sec:mot}). However, in our framework we chose to define labelled MESes
in terms of subsets of \emph{labels}, rather than subformulas. We argue that this 
definition is more natural. Consider, for example, the case of variable-MUSes (VMUSes).
Here, VMUS is a subset minimal set of \defterm{variables} of an unsatisfiable CNF formula,
rather than the subformula induced by these variables. If variables are used as labels
of clauses in the LCNF framework, as in Example~\ref{ex:2}(iii), then it is indeed the 
subset of labels of the formula that we are interested in, and not the subformula itself.% \anote{maybe another example}.
\begin{definition}[Labelled Minimal Equivalent Subset (LMES)]\label{def:lmes}
Let $\Phi = \tuple{\fml{F}, \lambda}$ be a labelled CNF formula. A set of labels 
$L \subseteq \lambda(\Phi)$ is a \defterm{labelled minimal equivalent subset (LMES)} of 
$\Phi$, if $\Phi|_L \equiv \Phi$, and $\forall L' \subset L$, $\Phi|_{L'} \notequiv \Phi$. 
The set of all LMESes of $\Phi$ is denoted by $\LMES(\Phi)$.
\end{definition}
As with (plain) CNF formulas, when $\Phi$ is unsatisfiable, LMESes of $\Phi$ capture the
generalized notion of minimally unsatisfiable subformulas.
\begin{definition}[Labelled Minimal Unsatisfiable Subset (LMUS)]\label{def:lmus}
Let $\Phi = \tuple{\fml{F}, \lambda}$ be a labelled CNF formula. A set of labels 
$L \subseteq \lambda(\Phi)$ is a \defterm{labelled minimal unsatisfiable subset (LMUS)} of 
$\Phi$, if $\Phi|_L \in \UNSAT$, and $\forall L' \subset L$, $\Phi|_{L'} \in \SAT$. The set 
of all LMUSes of $\Phi$ is denoted by $\LMUS(\Phi)$.
\end{definition}
To put the above definitions into a concrete context, consider the labelled CNFs 
discussed in Example~\ref{ex:2}: for the case $(i)$ the LMESes correspond to CNF-based 
MESes and LMUSes correspond to MUSes; for the case $(ii)$ the LMUSes correspond to
group-MUSes; for the case $(iii)$ the LMUSes correspond to variable-MUSes (VMUSes).
\begin{table}[t]
\caption{Summary of the corner cases for CNF and LCNF formulas. Here $\fml{F}$ refers to CNF formula, 
$\Phi$ to LCNF.}
\label{tbl:corner}
\centering
\begin{tabular}{|c||c|c|c|}
\hline
         & Exists for every formula ?  & Can be empty formula    & Can be the whole formula       \\
%         &                             & (resp. set of labels) ? & (resp. set of active labels) ? \\
\hline\hline
MES      & yes                        & yes, only when $\fml{F} = \emptyset$          & yes \\
\hdashline
LMES     & yes                        & yes, only when $\lambda(\Phi) = \emptyset$, or  & yes \\
         &                            & when $\fml{F}^\emptyset_\Phi \ne \emptyset$ and & \\
         &                            & all labels are redundant                      & \\
\hline
MNS      & no: when $\fml{F} = \emptyset$                & yes                                           & no \\
\hdashline
LMNS     & no: when $\lambda(\Phi) = \emptyset$, or      & yes                                           & no \\
         & when $\fml{F}^\emptyset_\Phi \ne \emptyset$ and &                                               &    \\
         & all labels are redundant                      &                                               &    \\
\hline
coMNS    & same as MNS                                   & no                                            & yes \\
\hdashline
coLMNS   & same as LMNS                                  & no                                            & yes \\
\hline
\end{tabular}
\vspace{-10pt}
\end{table}

Note that, by definition, when a label $l$ is irredundant in $\Phi$, \emph{every} 
LMES of $\Phi$ must include $l$, and, in fact, the set of all irredundant labels 
of $\Phi$ is precisely $\bigcap \LMES(\Phi)$. Thus, $\Phi$ is irredundant if and 
only if $\LMES(\Phi) = \{ \lambda(\Phi) \}$. Also, note that a label might be
redundant in $\Phi$, but irredundant in a subformula $\Phi'$ of $\Phi$. However,
if $l$ is irredundant in $\Phi$, it is irredundant in every subformula of $\Phi$.

Clearly, every labelled CNF formula $\Phi$ has at least one LMES, and, furthermore, 
for any subformula $\Phi'$ of $\Phi$, $\Phi' \equiv \Phi$ if and only if some 
LMES of $\Phi$ is a subset of $\lambda(\Phi')$. Note that in case of CNF formulas,
an MES can be empty only if the formula itself is empty. For the case of labelled
CNFs, an empty LMES can also occur when all labels are redundant --- but this 
can only happen in the presence of unlabelled clauses. Note that this additional 
case is not an artifact of the LCNF framework, but rather the artifact of the 
idea of group-0 clauses (in group-CNFs), and uninteresting variables (in the 
extensions of variable-MUSes). For example, group-MUS is empty when group-0 is 
unsatisfiable. Table~\ref{tbl:corner} contains a summary of this and other corner 
cases in the LCNF framework, and contrasts them with the corner cases in  (plain) 
CNF redundancy.

\begin{example}\label{ex:1b}
Consider the LCNF formula $\Phi$ from Example~\ref{ex:1}, for convenience we 
reproduce it here.
\begin{align*}
c_1 &= (\neg y)_{\{1\}}         & c_3 &= (z \lor t)_{\{1\}}   & c_5 &= (x \lor y \lor z)_\emptyset & c_7 &= (\neg y \lor t)_{\{3\}} \\
c_2 &= (y \lor \neg t)_{\{1\}}  & c_4 &= (\neg x)_{\{1,2\}}   & c_6 &= (\neg x \lor y)_{\{2,3\}}  & c_8 &= (\neg t)_{\{4\}}
\end{align*} 
To aid the understanding of the example note the following: the clauses $c_1, \dots, c_4$ are implied by 
the clauses $c_5, \dots, c_8$ ($c_1$ is derived from $c_7,c_8$ by resolution; $c_2$ is subsumed by $c_8$;
$c_3$ is derived from $c_5,c_6,c_7$; $c_4$ is derived from $c_6,c_7,c_8$); also, the clauses 
$c_6, c_7, c_8$ are implied by the clauses $c_1, c_2, c_4$ ($c_6$ is subsumed by $c_4$; $c_7$ is subsumed
by $c_1$; $c_8$ is derived from $c_1, c_2$).

Label 1 is redundant in $\Phi$ due to the fact that clauses $\fml{F}^1 = \{c_1,\dots,c_4\}$ are
implied by $\fml{F}|_{\{2,3,4\}} = \{ c_5,\dots,c_8 \}$. However, labels 2, 3 and 4 are irredundant
in $\Phi|_{\{2,3,4\}}$, hence $L_1 = \{2, 3, 4\}$ is a labelled MES of $\Phi$. 
The formula $\Phi$ has another LMES: label 3 is redundant in $\Phi$, as clauses $\fml{F}^3 = \{ c_6, c_7 \}$ are 
implied by $\fml{F}|_{\{1,2,4\}} = \{ c_1, \dots, c_5, c_8 \}$. However, $\Phi|_{\{1,2,4\}}$ contains a redundant 
label 4, as clause $c_8$ is implied by $c_1,c_2$. Now, $\Phi|_{\{1,2\}} = \tuple{\{ c_1, \dots, c_5 \}, \lambda}$ is 
irredundant --- even though clause $c_5$ is implied by $c_2$ and $c_3$ and so is redundant in the (plain) CNF
sense, we cannot remove it from $\Phi|_{\{1,2\}}$; note that this would also be the case if $\lambda(c_5) = \{ 2 \}$.
We conclude that $L_2 = \{ 1, 2\}$ is an LMES of $\Phi$.
\end{example}

The notion dual to minimal equivalence (resp. minimal unsatisfiability) is that of maximal
non-equivalence (resp. maximal satisfiability). 
%Maximal satisfiability is of particular interest
%to AI due to the applications to reasoning in the presence of inconsistency (for example, \cite{lozinsikii}).
Here we are interested in sets of labels that induce a subformula of $\Phi$ that is not equivalent 
to $\Phi$, but an addition of any active label from $\Phi$, results in an equivalent subformula.
\begin{definition}[Labelled Maximal Non-equivalent Subset (LMNS)]\label{def:lmns}
Let $\Phi = \tuple{\fml{F}, \lambda}$ be a labelled CNF formula. A set of labels 
$L \subseteq \lambda(\Phi)$ is a \defterm{labelled maximal non-equivalent subset (LMNS)} of 
$\Phi$, if $\Phi|_L \notequiv \Phi$ and for every $L'$, $L \subset L' \subseteq \lambda(\Phi)$,
$\Phi|_{L'} \equiv \Phi$. The set of all LMNSes of $\Phi$ is denoted by $\LMNS(\Phi)$.
\end{definition}
Note that just as with clausal MNSes, which do not exist for empty formulas because every
subformula of an empty formula is equivalent to it, LMNSes do not exist for LCNF formulas with 
$\lambda(\Phi) = \emptyset$. Also, just as with LMESes, the presence
of unlabelled clauses gives rise to an additional corner case (see also Table~\ref{tbl:corner}) --- 
when all labels are redundant (for non-empty formulas this can only happen if 
$\fml{F}^\emptyset \ne \emptyset$), every subformula of $\Phi$ is also equivalent to $\Phi$.  
For the case of unsatisfiable LCNFs, we have a definition analogous to that of (clausal) MSS.
\begin{definition}[Labelled Maximal Satisfiable Subset (LMSS)]\label{def:lmss}
Let $\Phi = \tuple{\fml{F}, \lambda}$ be a labelled CNF formula. A set of labels 
$L \subseteq \lambda(\Phi)$ is a \defterm{labelled maximal satisfiable subset (LMSS)} of 
$\Phi$, if $\Phi|_L \in \SAT$ and for every $L'$, $L \subset L' \subseteq \lambda(\Phi)$,
$\Phi|_{L'} \in \UNSAT$. The set of all LMSSes of $\Phi$ is denoted by $\LMSS(\Phi)$.
\end{definition}
Note that as opposed to MSSes, which exist for every CNF formula, LMSSes do not exist for 
formulas with an unsatisfiable set of unlabelled clauses, because no subformula of such
a formula is satisfiable. 

As discussed in Section~\ref{sec:mot}, clausal MSSes are of interest for a number of 
reasons, one of which that an MSS of maximum cardinality is a set of clauses that are 
true under a solution to MaxSAT problem. With this in mind we can also define a generalized 
version of MaxSAT problem.

%Traditionally, MSSes were used to allow
%reasoning in the presence of inconsistency in inconsistent knowledge bases. Here an MSS represents
%a maximally consistent part of the knowledge base, which can be used to derive new facts. 
%While in many earlier approaches a sinlge MSS wad use for this purpose (e.g. one of maximal cardinality,
%see below), a number of researches, for example \cite{lozinskii-jetai03}, have proposed to use a set of all MSSes, 
%or certain subsets of it. 
%\anote{TODO: define LMaxSAT}

Given an LMSS $L$ of $\Phi$, one may also consider its complement 
$\lambda(\Phi) \setminus L$. When $\Phi \in \SAT$, the complement is an empty set, 
however when $\Phi \in \UNSAT$, $\lambda(\Phi) \setminus L$ is a minimal 
set of labels of $\Phi$, removal of which from $\Phi$, will regain the satisfiability. 
The corresponding concept in the context of unsatisfiable CNF is that of co-MSS (cf. Section~\ref{sec:mot}).
% co-MSSes have
%been under the radar of AI communitiy since the early work of Reiter \cite{reiter} on diagnosis where they 
%were referred to as minimal diagnoses.\anote{edit}
%
Similar, though less intuitive, concept arises in the case of LMNSes.
\begin{definition}[co-LMNS]\label{def:colmns}
Let $\Phi = \tuple{\fml{F}, \lambda}$ be a labelled CNF formula. A set of labels 
$L \subseteq \lambda(\Phi)$ is a \defterm{labelled co-MNS (co-LMNS)} of $\Phi$, if 
$\lambda(\Phi) \setminus L \in \LMNS(\Phi)$. Or, explicitly, if
$\Phi_{\lambda(\Phi) \setminus L} \notequiv \Phi$, and for any $L' \subset L$, 
$\Phi_{\lambda(\Phi) \setminus L'} \equiv \Phi$. The set of all co-LMNSes of $\Phi$ is 
denoted by $\coLMNS(\Phi)$.
\end{definition}
\begin{definition}[co-LMSS]\label{def:colmss}
Let $\Phi = \tuple{\fml{F}, \lambda}$ be a labelled CNF formula. A set of labels 
$L \subseteq \lambda(\Phi)$ is a \defterm{labelled co-MSS (co-LMSS)} of $\Phi$, if 
$\lambda(\Phi) \setminus L \in \LMSS(\Phi)$. Or, explicitly, if
$\Phi_{\lambda(\Phi) \setminus L} \in \SAT$, and for any $L' \subset L$, 
$\Phi_{\lambda(\Phi) \setminus L'} \in \UNSAT$. The set of all co-LMSSes of $\Phi$ is 
denoted by $\coLMSS(\Phi)$.
\end{definition}
\begin{example}\label{ex:1c}
Consider again the LCNF formula $\Phi$ from Example~\ref{ex:1}. The formula has three LMNSes: 
$\{ 1, 3, 4 \}$, $\{ 2, 3 \}$ and $\{2, 4\}$, and three corresponding co-LMNSes.
\end{example}
%
% table was here

\subsection{Generalized Hitting Set Duality}

As mentioned in Section~\ref{sec:mot}, for a given CNF formula $\fml{F}$, there is a 
relationship between the set of MUSes of $\fml{F}$ and the set of co-MSSes of 
$\fml{F}$: $\coMSS(\fml{F})$ is a set of irreducible hitting sets of $\MUS(\fml{F})$.
This relationship has been (re)discovered on a 
number of occasions, with the earliest,
to our knowledge, attributed to Reiter \cite{reiter-aij87} in the context of model-based
diagnosis --- there MUSes are called minimal conflict sets, and coMSSes are called 
minimal diagnoses. This relationship is a basis for the efficient MUS enumeration
algorithms (cf. \cite{stuckey-padl05,kas-jar08}. A weaker form of this relationship, namely
$\bigcup \MUS(\fml{F}) = \fml{F} \setminus \bigcap \MSS(\fml{F})$, derived by
Kullmann \cite{kullmann-dam03}, has been also generalized in \cite{kullmann-fi11b} 
to the case of satisfiable CNF formulas. In this section we develop a general version
of the hitting set theorem for the labelled CNF formulas. In addition to subsuming
the previous results, the theorem covers all the other, not previously analyzed,
cases, e.g. group-MUS or variable-MUS. The theorem also allows to develop effective algorithm
computation of the set of all LMESes.

The proof of the theorem relies on a number of basic properties of LMESes and LMNSes,
as well as the following known property of irreducible hitting sets (recall Definition~\ref{def:hs}).
The property asserts that every element of an irreducible hitting set must, in a sense, have 
a ``reason'' to be there, i.e. to be a unique representative of some set.

\begin{proposition}\label{pr:hs}
Let $\class{S}$ be a collection of arbitrary sets, and let $H$ be any hitting set of $\class{S}$.
Then, $H$ is irreducible if and only if $\forall h \in H$, $\exists S \in \class{S}$ such that
$H \cap S = \{ h \}$. 
\end{proposition}
%We omit the proof of Proposition~\ref{pr:hs}. 
The hitting sets relationship is captured formally by the following theorem.
\begin{theorem}[Generalized Hitting Set Duality Theorem]\label{th:duality}
Let $\Phi = \tuple{\fml{F}, \lambda}$ be a labelled CNF formula, such that $\lambda(\Phi) \ne \emptyset$,
and if $\fml{F}^\emptyset \ne \emptyset$ then at least one label in $\lambda(\Phi)$ is irredundant.
Then,
\begin{itemize}
\item[(i)] $L \subseteq \lambda(\Phi)$ is a coLMNS of $\Phi$ if and only if $L$ is an irreducible
hitting set of $\LMES(\Phi)$.
\item[(ii)] $L \subseteq \lambda(\Phi)$ is an LMES of $\Phi$ if and only if $L$ is an irreducible 
hitting set of $\coLMNS(\Phi)$.
\end{itemize}
\vspace{-10pt}
\end{theorem}

Note that the restrictions on the formula $\Phi$ in the above theorem are in place to ensure
that the formula has at least one co-LMNS (cf. Table~\ref{tbl:corner}). These restrictions 
are satisfied \emph{a priori} for a number of special cases, which we discuss shortly.

The intuition behind \emph{(i)} can be explained as follows\footnote{This explanation is
a generalized version of the one given for unsatisfiable CNF case in \cite{kas-jar08}} --- 
since the removal of a co-LMNS from a formula $\Phi$ makes it non-equivalent to $\Phi$, the 
removal must ``break'' each of the LMESes of the formula. Hence a co-LMNS must include at 
least one label from each of the LMESes, i.e. it is a hitting set of the set of LMESes of 
the formula. The minimality of co-LMNS implies the irreducibility of the hitting set, and 
vice versa.

Before we proceed with the proof of Theorem~\ref{th:duality}, recall a simple property
of subformulas of any LCNF formula $\Phi$ that satisfies the conditions of the theorem:
%\anote{make sure these are stated somewhere earlier}: 
for any $\Phi' \subseteq \Phi$, 
$\Phi' \notequiv \Phi$ if and only if $\lambda(\Phi')$ is a subset of some LMNS of 
$\Phi$; $\Phi' \equiv \Phi$ if and only if $\lambda(\Phi')$ is a superset of some
LMES of $\Phi$.

\begin{proof} For clarity we adopt the following convention: letter $S$ will be used
to denote LMNSes, $M$ to denote co-LMNSes, $U$ to denote LMESes.

%\begin{itemize}
%\item[(i)]
\emph{Part (i), If:} Let $M$ be an irreducible hitting set of $\LMES(\Phi)$, and let
$S = \lambda(\Phi) \setminus M$. First, since $M$ is a hitting set of $\LMES(\Phi)$,
$S$ cannot include an LMES of $\Phi$, and so $\Phi|_S \notequiv \Phi$.
Since $M$ is an \emph{irreducible} hitting set of $\LMES(\Phi)$, for any label $l \in M$,
there exists $U \in \LMES(\Phi)$, such that $M \cap U = \{ l \}$ (by Proposition~\ref{pr:hs}).
Hence, for any $l \in M$, the set $S \cup \{ l \}$ includes some LMES $U$ of 
$\Phi$, and so $\Phi|_{S \cup \{ l \}} \equiv \Phi$. We conclude that $S$ is an LMNS of 
$\Phi$, and so $M$ is a co-LMNS of $\Phi$.

\emph{Part (i), Only-if:} Let $M$ be any co-LMNS of $\Phi$, and let $S = \lambda(\Phi) \setminus M$
be the corresponding LMNS. Since $\Phi|_S \notequiv \Phi$, for any $U \in \LMES(\Phi)$, 
$U \setminus S \ne \emptyset$ (otherwise $U \subseteq S$), and so
$U \cap M \ne \emptyset$, that is, $M$ is a hitting set of $\LMES(\Phi)$.
Now, since $S$ is an LMNS, for every label $l \in M$, $\Phi|_{S \cup \{ l \}} \equiv \Phi$.
Thus, for every $l \in M$, there exists an LMES $U$ such that $M \cap U = \{ l \}$.
By Proposition~\ref{pr:hs}, $M$ is an \emph{irreducible} hitting set of $\LMES(\Phi)$.

%\item[(ii)] 
\emph{Part (ii), If:} Let $U$ be an irreducible hitting set of $\coLMNS(\Phi)$. We have 
that for any $M \in \coLMNS(\Phi)$, $U \cap M \ne \emptyset$. Hence, for no
$S \in \LMNS(\Phi)$ we have $U \subseteq S$ and so $\Phi|_U \equiv \Phi$. Since
$U$ is \emph{irreducible}, by Proposition~\ref{pr:hs}, for every label $l \in U$, there
exists $M \in \coLMNS(\Phi)$ such that $U \cap M = \{ l \}$. Thus, for
every $l \in U$, there exists a co-LMNS $M$ such that 
$U' = U \setminus \{ l \} \subseteq \lambda(\Phi) \setminus M$, i.e. $U'$
is included in some LMNS of $\Phi$, and so $\Phi|_{U'} \notequiv \Phi$. We conclude that 
$U \in \LMES(\Phi)$.

\emph{Part (ii), Only-if:} Let $U$ be any LMES of $\Phi$. Since $\Phi|_U \equiv \Phi$, $U$
cannot be included in any LMNS of $\Phi$, and so for every co-LMNS $M$ of $\Phi$,
we have $U \cap M \ne \emptyset$, i.e. $U$ is a hitting set of $\coLMNS(\Phi)$.
Now, since $U$ is an LMES of $\Phi$, for any label $l \in U$, $\Phi|_{U \setminus \{ l \}} \notequiv \Phi$,
and so the set $U \setminus \{ l \}$ is included in some LMNS of $\Phi$. Hence, for any label $l \in U$, 
there exists a co-LMNS $M$ of $\Phi$ such that $U \cap M = \{ l \}$. Hence, 
By Proposition~\ref{pr:hs}, $U$ is an \emph{irreducible} hitting set of $\coLMNS(\Phi)$.
%\end{itemize}
\qed
\end{proof}

The restrictions on the formula $\Phi$ in Theorem~\ref{th:duality} can, in some cases,
be satisfied \emph{a priori}. Consider, for example, the case $\Phi \in \UNSAT$, and the labelling
function as in Example~\ref{ex:2}(i). Since $\fml{F}_\Phi \in \UNSAT$, we have 
$\fml{F} \ne \emptyset$, and every clause is labelled ($\fml{F}^\emptyset = \emptyset$),
the theorem applies unconditionally to such formulas. Thus, we get exactly the
original version of hitting set duality theorem for unsatisfiable CNF formulas (see Section~\ref{sec:mot}).
For the case of group-MUS (Example~\ref{ex:2}(ii)), the theorem holds whenever 
$\fml{F}^\emptyset \in \SAT$, as this condition ensures that
the formula has at least one irredundant label (since $\Phi \in \UNSAT$). 

The following corollary is a straightforward consequence of Theorem~\ref{th:duality},
and is a generalized version of the relationship between MUSes and
co-MSSes shown in \cite{kullmann-cc00}. % kullmann-dam03,kullmann-fi11b}.
\begin{corollary}\label{col:kul}
Let $\Phi$ be as in Theorem~\ref{th:duality}. Then, $\bigcup \LMES(\Phi) = \lambda(\Phi) \setminus \bigcap \LMNS(\Phi)$.
\end{corollary}
The following example illustrates the claims of Theorem~\ref{th:duality} and Corollary~\ref{col:kul}.
\begin{example}\label{ex:1c}
Consider the LCNF formula $\Phi$ from Example~\ref{ex:1}. From Examples~\ref{ex:1b} and
\ref{ex:1b} we have the following: $\LMES(\Phi) = \{ \{ 1, 2 \}, \{ 2, 3, 4 \} \}$, 
$\LMNS(\Phi) = \{ \{ 1, 3, 4 \}, \{ 2, 3 \}, \{2, 4\} \}$, 
$\coLMNS(\Phi) = \{ \{ 2 \}, \{ 1, 3 \}, \{ 1, 4 \} \}$. Note that $\LMES(\Phi)$ has 
exactly 3 irreducible hitting sets that constitute the set $\coLMNS(\Phi)$. Also,
$\bigcup \LMES(\Phi) = \{ 1, 2, 3, 4 \} = \lambda(\Phi)$, and $\bigcap \LMNS(\Phi) = \emptyset$.
\end{example}

\iffalse % TODO: figure this out properly

\subsection{A Note on Complexity}

Labelled CNF formulas appear to be rather uninteresting from the computational complexity
point of view. In fact, since the labelling function is a part of the formula,
the description of a labelled CNF formula can be exponential in the number of clauses in the
CNF part of the formula, even if no duplicate labels (i.e. those labels that mark 
exactly the same sets of clauses) are present. For such formulas SAT decision is
in $\PP$, since the complexity of SAT decision depends only on the size of the CNF part 
of the formula. 
For LCNF formulas whose size is polynomial in the size of their CNF part, the labelling
does not affect any of the complexity results for (plain) CNF formulas.
\fi

%%% Local Variables:
%%% mode: pdflatex
%%% TeX-master: "paper"
%%% End:

%\input{algorithms}
%
% Generalized redundancy (arXiv version)
%

%\paragraph{\bf Conclusion}\label{sec:conc}
\section{Conclusion}\label{sec:conc}

This report presents a framework of labelled CNF formulas that allows to generalize
and extend the existing work on redundancy detection and removal in CNF formulas. 
Future work includes the development of a number of additional theoretical results,
and a suite of efficient algorithms that address various computational problems in 
the context of the proposed framework.

%%% Local Variables:
%%% mode: pdflatex
%%% TeX-master: "paper"
%%% End:

%\input{scratchpad}

%\bibliographystyle{plain}
\bibliographystyle{abbrv}
\bibliography{minunsat,refs,paper}

\end{document}